\documentclass[dvips]{article}
\usepackage{icrctc07}

\title{$\gamma$-ray emission from PWNe interacting with molecular clouds}
\shorttitle{$\gamma$-ray emission from PWNe}

\authors{Hendrik Bartko $^{1}$, W\l odek Bednarek$^{2}$}
\shortauthors{Author and et al.}
\afiliations{$^1$Max Planck Institut f\"ur Physik, M\"unchen, Germany,\\ $^2$Department of Experimental Physics, University of \L \'od\'z, \L \'od\'z, Poland }
\email{hbartko@mppmu.mpg.de; bednar@fizwe4.phys.uni.lodz.pl}

\abstract{We consider a situation in which a pulsar (and its nebula) is formed inside or close to a high density regions of a molecular cloud. We apply a recent model for the gamma radiation of 
pulsar wind nebulae (PWN), which includes not only radiation processes due to injected leptons but also processes due to injection of relativistic hadrons, in order to calculate the expected 
$\gamma$-ray emission from such interacting PWNe. The example calculations have been performed for two objects of this type from which directions TeV $\gamma$-ray sources have been recently observed (IC443 and W41). We show that the $\gamma$-ray emission below a few TeV can be produced by leptons accelerated in the past in the vicinity of the pulsars. $\gamma$-rays with energies above $\sim 10$ TeV can be produced by hadrons
interacting with the matter inside the supernova remnant and surrounding dense clouds.
In contrary to the low energy TeV emission, this high energy TeV emission should be correlated with the location of dense clouds able to capture hadrons due to their strong magnetic fields.
}

\begin{document}
\maketitle
\section{Introduction}

A few recently discovered TeV $\gamma$-ray sources seem to be clearly related to Pulsar Wind Nebulae (PWNe)~\cite{ah06}. One of the most interesting cases is the Vela pulsar and its nebula. This is the closest object of this type at a distance of only $\sim 300$ pc. However, the center of the TeV source is clearly displaced from the present position of the pulsar by $\sim 0.5^{\rm 0}$. This displacement can be explained by the fact that the emission could come from particles injected at the early age of the pulsar and the pulsar moved significantly during its age ($\sim$10$^4$ yrs) from its position at birth. Here, we consider in more detail two such example supernova remnants with evidences of associated energetic pulsars: IC 443, recently discovered by the MAGIC telescope (MAGIC J0616+225,~\cite{al07}) and W41, discovered by the HESS telescopes (HESS J1834-087,~\cite{ah06}) and confirmed by the MAGIC telescope~\cite{al06}. 

The spectrum of the TeV source towards IC 443 is relatively steep (spectral index $-3.1$). The source is located in the direction of dense molecular clouds which probably lay in front
of this asymmetric SNR. Moreover, an X-ray nebula has been discovered in this direction by recent {\it XMM} and {\it Chandra} observations~\cite{bb01,ol01}. It is being interpreted as due to the presence of an energetic Vela type pulsar. The X-ray PWNa is displaced from the location of the TeV source by $\sim 20'$. We suggest that this pulsar can also be responsible for the observed TeV $\gamma$-ray source. $\gamma$-rays can be produced by particles injected from the PWNa during the early age of the pulsar. In fact, the pulsar with a velocity of the Vela pulsar ($\sim$ 250 km s$^{-1}$) should change its position by $\sim 18'$ during the age of $3\times 10^4$ yrs. The distance crossed by such a pulsar is very similar to the displacement between the position of the observed PWNa and the MAGIC TeV source.

The TeV spectrum of the second source (HESS J1834-087) is flatter, spectral index $\sim 2.5$. The source looks extended with a size of $\sim 12'$~\cite{ah06,al06}. It is positionally coincident with the shell-type SNR, G23.3-0.3 (W41), with an age estimated as $\sim 8\times 10^4$ yrs. A Vela type pulsar, with the period of 85 ms, is also possibly connected with this supernova~\cite{ga95}. Its characteristic age and distance are consistent with the age and distance of W41. However, the pulsar is displaced from the center of the SNR (and also from the position of the TeV source) by $\sim 24'$. Such positional disagreement might be explained by the movement of the pulsar during $\sim 10^5$ yrs with a velocity of $\sim 250$ km s$^{-1}$. The pulsar seems to be located in the direction of one of the extensions of the TeV $\gamma$-ray source~\cite{ah06}.

\section{A model for $\gamma$-ray production}

We consider a scenario in which an energetic, fast moving, pulsar is created in a supernova explosion  close to a dense cloud. Such a picture can differ significantly from an isolated pulsar wind nebula model since the huge nearby concentration of matter can provide additional soft radiation and matter targets for particles accelerated in the vicinity of the pulsar. Moreover, the pulsar, changing position in time, injects particles (leptons, hadrons) in different places. This situation provides a unique opportunity for studying the time dependence of the high energy processes around pulsars.

We calculate the $\gamma$-ray spectra from PWNe in the vicinity of dense clouds by
applying the hybrid (leptonic-hadronic) time dependent model (details in~\cite{bb03}). In summary, the model applies the hypothesis of Arons and collaborators~\cite{ho92,ga94}, according to which leptons gain energy by being resonantly scattered on the Alfven waves generated above the pulsar wind shock by hadrons injected by the pulsar. This model assumes that most of the rotational energy loss rate of the pulsar is taken by relativistic hadrons which can gain energies corresponding to up to $20\%$ of the maximum potential drop through the pulsar magnetosphere. In the time dependent model for the PWNe considered in~\cite{bb03}, the lower energy TeV $\gamma$-ray emission (below a few TeV) is mainly produced in the ICS process by leptons accelerated in the pulsar nebula and the higher TeV emission by hadrons inside the nebula. The highest energy $\gamma$-rays ($10^2-10^3$ TeV) are produced
in a more extended region by the highest energy hadrons which escaped from the nebula in the past and were captured inside dense nearby clouds. Therefore, in principle, TeV emission below a few TeV should be related to the volume of the nebula itself but the highest energy emission should be correlated with the distribution of dense clouds surrounding the nebula.

The analysis of the radiation processes around pulsars is further complicated in the case of pulsars 
which are able to significantly change their position in  respect to the birth place. 
This situation obviously concerns the Vela type pulsars which are still
energetic enough to produce magnetospheric $\gamma$-rays and reached large velocities at birth.
In their case, radiation processes can be additionally distributed in space.

\section{IC 443/MAGIC J0616+225}

The parameters of the pulsar which creates the PWNa towards IC 443 are unknown.
But from comparison with other nebulae, it is argued that the pulsar has to be 
a Vela type pulsar with a present period of $\sim$145 ms, a surface 
magnetic field $\sim$3$\times 10^{12}$ G, and an age of 
$\sim$3$\times 10^4$ yrs~\cite{ol01}. The displacement of the 
PWNa in respect to the center of the SNR can be explained by the motion of the pulsar with a velocity of $\sim$250 km s$^{-1}$ and applying the age of the SNR.
We assume that the supernova exploded in the medium with a density of $\sim 20$
particles / cm$^{3}$ and a magnetic field of $10^{-5}$ G. Such a large density is 
supported by the presence of dense clouds around the SNR.
The pulsar with the parameters mentioned above resembles the Vela pulsar and 
PSR 1706-44. 
In Fig. 1, we compare the $\gamma$-ray spectrum observed from PSR 1706-44
with the spectrum of the EGRET source 3EG J0617+2238 \cite{EGRET} which is observed in
the direction of IC 443. There is good consistency between the
fluxes and spectra of these two sources. Therefore, we propose that,
in spite of only marginal directional agreement, the EGRET source 3EG J0617+2238 is due to the pulsar responsible for the PWNa observed towards IC 443.

Applying the model for the time evolution of high energy processes inside 
PWNe~\cite{bb03}, we calculate the $\gamma$-ray 
spectra produced by leptons in the IC and bremsstrahlung processes in case of
a nebula with the parameters of the pulsar and surrounding medium mentioned above assuming  the age of the pulsar and SNR to be equal to $3\times 10^4$ yrs. The initial period of the pulsar is assumed to be equal to 10 ms. The difference between the present location of the PWNa
and the position of the MAGIC TeV $\gamma$-ray source can be explained by 
assuming that leptons are accelerated by hadrons only during the early
phase of the PWNa. In such cases, leptons were accelerated in the past at a 
different place than the present location of the pulsar. We assume that this initial
activity stage of lepton acceleration is limited to $10^4$ yrs, i.e.
the age of the present Vela pulsar. However, since these  leptons were
injected close to the dense regions of the cloud, the infrared soft 
radiation field there is expected to be considerably stronger due to the 
emission from the dust heated by the supernova shock. 
Following other works, we apply the soft radiation field 
of the interstellar medium at the galactic disk~\cite{de97,ga98}. It is composed of the cosmic microwave background radiation (MBR) with a temperature of 2.7 K, the infrared background with energy 
density 2 times larger than the MBR and a temperature of 25 K, the optical 
background defined by energy densities equal to the MBR and with characteristic temperatures between 5000 K and $10^4$ K. Moreover, we add the infrared radiation field with an energy density 3 times larger than the MBR and with a temperature of 45 K, which is supposed to be produced by dust (based on the IRAS data, Mufson et al. 1986). The $\gamma$-ray spectra produced by leptons 
in the IC and bremsstrahlung processes are shown in Fig. 1. We assume that
hadrons take $95\%$ of the rotational energy loss of the pulsar and $5\%$ of the energy of hadrons is transfered to leptons with a power law spectrum and a differential spectral index equal to 2.4.

\begin{figure*}[t]
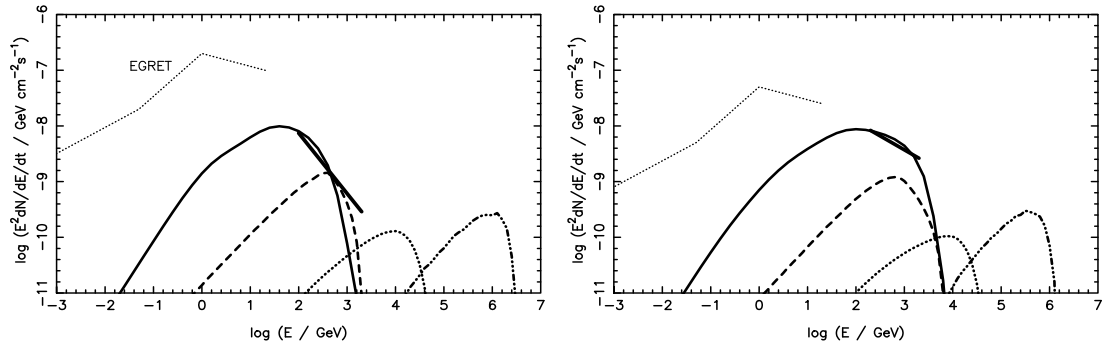

\vskip 4.5truecm
\includegraphics{icrc91_fig1a.eps}
\includegraphics{icrc91_fig1b.eps}
\caption{\label{fig1} The $\gamma$-ray spectra (SED) produced by leptons in the
IC and bremsstrahlung processes inside the PWN with the age of $3\times 10^4$ yrs
(in the case of IC 443, left figure) and $8\times 10^4$ yrs (W41, right figure).
Leptons are injected during the first $10^4$ yrs (IC 443) and $4\times 10^4$ yrs (W41) 
after the supernova explosion (solid and 
dashed curves, respectively), the $\gamma$-ray spectra produced by hadrons 
inside the supernova remnant (thick dotted), and by hadrons which escaped 
from the supernova remnant and were captured by nearby dense clouds 
(triple-dot-dashed curves). We also show the EGRET $\gamma$-ray spectrum observed from the inner 
magnetosphere of the Vela type pulsar PSR 1706-44, assuming that it is at 
the distance of the pulsar related to these TeV sources (thin dotted).
The density of matter inside the supernova was assumed to be 20 cm$^{-3}$ and in the molecular 
clouds $10^3$ cm$^{-3}$. We assumed that $10^{-5}$ of escaping hadrons are captured 
by dense clouds. The other parameters of the model are given in the text. The thick lines mark
the TeV spectrum observed by the MAGIC telescope.}
\end{figure*}

We also calculate the $\gamma$-ray spectrum from decays of $\pi^{\rm o}$
which are produced by hadrons inside the nebula. In the calculations of the $\gamma$-ray 
spectra we apply the scaling break model for hadronic interactions~\cite{wd87}. However, since the nebula is relatively old, most of the hadrons 
accelerated at the early stage of the nebula have already escaped from it. 
Therefore, the level of the $\gamma$-ray emission produced by hadrons which are 
still inside the SNR is relatively low (see Fig. 1).

We also estimate the $\gamma$-ray fluxes produced by hadrons which escaped
from the PWNa in the past activity stage of the pulsar. It is very difficult to reliably estimate the part of hadrons which are captured inside dense clouds, since this process depends on the geometry of the cloud and nebula, their relative distance, and after 
all on the geometry and distribution of the magnetic field in the region surrounding the PWNa.
This accumulation of hadrons may (or may not) also depend on the energy of the accelerated hadrons, depending on the details of the capturing process (e.g. whether it is mainly due to energy 
dependent diffusion or due to advection with the wind from the pulsar). 
Since we are not able to take into account these complicated processes, it is assumed 
that a part of all hadrons escaping from the PWNa is captured inside dense clouds.
The example calculations are performed for the parameters of the PWNa mentioned above, 
assuming the presence of a nearby cloud with a density of $10^3$ particles cm$^{-3}$
and efficiency of accumulation of hadrons equal to $10^{-5}$. 
It is clear, that even for such a low capturing efficiency, these hadrons can produce
large fluxes of $\gamma$-rays at energies $\sim$10-10$^3$ TeV. 
Note that this high energy component in the $\gamma$-ray spectrum should follow the 
distribution of dense matter around the PWNa. Its investigation should provide 
important additional constraints on the evolution of the supernova remnant.
The level of this emission can also be constrained by the future 1 km$^3$ neutrino
detectors whose sensitivity peaks in this energy range.

\section{W41/PSR J1833-827/HESS J1834-087}

The parameters of the pulsar, PSR J1833-827, are known. The distance, $\sim$4 kpc, and the characteristic age of the pulsar, $\sim 10^5$ years, are consistent with the corresponding parameters of the SNR G23.3-0.3 (W41, i.e. an age of $8\times 10^4$ yrs~\cite{ti06}). As in the case of IC 443, W41 is associated with a large molecular complex called "[23,78]~\cite{da86} at similar distance. W41 seems to be another good candidate for the interaction of particles accelerated by a pulsar with dense matter and an infrared radiation field produced by dust. Therefore, we also apply for this object the time dependent model for the acceleration of hadrons and leptons in the vicinity of the energetic pulsar. The general parameters of the medium with which particles interact, are kept the same, as in the case of IC 443. The initial period of the pulsar is assumed as before, i.e. 10 ms, and the surface magnetic field is chosen in such a way as to get a correct value for the present pulsar period, i.e. 85 ms. The value of the magnetic field ($1.2\times 10^{12}$ G) is close to the one estimated from the known pulsar period and the period derivative, and applying the pulsar rotating magnetic dipole model. The results of calculations of the IC and bremsstrahlung spectra produced
by leptons, which are injected during the first $4\times 10^4$ years after the pulsar formation, are shown in Fig. 1. Note, that these leptons produce $\gamma$-rays at the past position of the pulsar but not at its present position. We also calculate the $\gamma$-ray spectra from
decay of pions produced by hadrons, accelerated in the vicinity of the pulsar, which are
still inside the SNR, and those ones which escaped from the SNR and interact with dense clouds.
$\gamma$-rays produced by hadrons have typically larger energies than those ones produced by leptons. This is due to the large energy losses of leptons injected in the past. 
The highest energy $\gamma$-rays ($10^2-10^3$ TeV) are produced by hadrons
which escaped from the SNR and are captured by strong magnetic fields of the molecular clouds.
Therefore, this highest energy $\gamma$-ray emission should have a different distribution than the lower $\gamma$-ray emission, below a few TeV. 
As in the case of IC 443, this emission should be strongly correlated with dense regions of the molecular clouds. In contrast, lower energy $\gamma$-ray emission should be related to the birth place of the pulsar, i.e. the center of the SNR.


\begin{thebibliography}{}
\bibitem{ah06} Aharonian, F.~A. et al., 2006, ApJ, 636, 777.
\bibitem{al06} Albert, J. et al., 2006, ApJ, 643, L53.
\bibitem{al07} Albert, J. et al. 2007, submitted to ApJ, arXiv:0705.3119. 
\bibitem{bb03} Bednarek, W., Bartosik, M., 2003, A\&A 405, 689.
\bibitem{bb01} Bocchino, F., Bykov, A., 2001, A\&A 376, 248.
\bibitem{da86} Dame, T.~M. et al., 1986, ApJ, 305, 892.
\bibitem{de97} De Jager, O.~C., Mastichiadis, A., 1997, ApJ, 482, 874.
\bibitem{ga98} Gaisser, T.~K., Protheore, R.~J., Stanev, T., 1998, ApJ, 492, 219.
\bibitem{ga94} Gallant, Y.~A., Arons, J., 1994, ApJ, 435, 230.
\bibitem{ga95} Gaensler, B.~M., Johnston S., 1995, MNRAS, 275, L73.
\bibitem{EGRET} Hartman, R.~C. et al., 1999, ApJS, 123, 79.	
\bibitem{ho92} Hoshino, M. et al., 1992, ApJ, 390, 454.
\bibitem{mu86} Mufson, S.~L. et al., 1986, AJ, 92, 1349.
\bibitem{ol01} Olbert, C.~M. et al., 2001, ApJ, 554, L205.
\bibitem{ti06} Tian, W.~W. et al., 2007, ApJ, 657, L25.
\bibitem{wd87} Wdowczyk, J., Wolfendale, A.~W., 1987, J.Phys. G, 13, 411.

\end{thebibliography}
\end{document}